\documentclass[a4paper,11pt]{article}
\usepackage{amsmath, amssymb, authblk, a4wide, cite, graphicx, hyperref}

\newcommand{\pfrac}[2]{\left(\frac{#1}{#2}\right)}

\def\be{\begin{equation}}
\def\ee{\end{equation}}
\def\ba{\begin{eqnarray}}
\def\ea{\end{eqnarray}}

\def\uno{\mbox{1 \kern-.59em {\rm l}}}

\renewcommand\Affilfont{\itshape\small}

\begin{document}

\title{\Large{\textbf{Natural gauge mediation with a bino NLSP at the LHC}}}
\author[1]{James Barnard}
\author[2]{Benjamin Farmer}
\author[1,3]{Tony Gherghetta}
\author[1]{Martin White}
\affil[1]{ARC Centre of Excellence for Particle Physics at the Terascale,
School of Physics, University of Melbourne, Victoria 3010, Australia}
\affil[2]{ARC Centre of Excellence for Particle Physics at the Terascale, School of Physics, Monash University, Victoria 3800
Australia}
\affil[3]{Stanford Institute of Theoretical Physics, Stanford University, Stanford, CA 94305, USA}
\date{}
\maketitle

\begin{abstract}
\baselineskip=15pt
\noindent Natural models of supersymmetry with a gravitino LSP provide distinctive signatures at the LHC\@.  For a neutralino NLSP, sparticles can decay to two high energy photons plus missing energy.  We use the ATLAS diphoton search with 4.8 fb$^{-1}$ of data to place limits in both the stop-gluino and neutralino-chargino mass planes for this scenario.  If the neutralino is heavier than 50 GeV, the lightest stop must be heavier than 580 GeV, the gluino heavier than 1100 GeV\@ and charginos must be heavier than approximately 300-470 GeV\@.  This provides the first nontrivial constraints in natural gauge mediation models with a neutralino NLSP decaying to photons, and implies a fine tuning of at least a few percent in such models.
\end{abstract}
\maketitle
\section{Introduction}

Supersymmetry remains the most favourable solution to the hierarchy problem.  Not only does it elegantly cancel the quadratic divergences of an elementary Higgs boson, but it also achieves a precise gauge coupling unification without the need for large threshold effects.  The Large Hadron Collider (LHC) has begun to systematically exclude a sizeable portion of the parameter space in the supersymmetric standard model.  In particular, the limits on squark and gluino masses suggest that minimal versions of the supersymmetric standard model have become increasingly tuned, putting into question the original motivation for weak scale supersymmetry.

However, tuning in the Higgs sector of the supersymmetric standard model depends mainly on the $\mu$-term and on sparticles with large couplings to the Higgs boson, namely the third generation sfermions, gluinos and electroweak gauginos.  If the supersymmetry breaking mechanism is actually flavour dependent, third generation sfermions can be much lighter than the first two generations.  The scenario has recently been coined ``natural supersymmetry" \cite{Dimopoulos:1995mi, Pomarol:1995xc, Cohen:1996vb, Chou:1999zb, Asano:2010ut, Essig:2011qg, Brust:2011tb, Papucci:2011wy} and, together with the fact that the stop production cross section is much smaller than that of the up and down squarks at the LHC, it allows the stringent limits on squark and gluino masses to be alleviated.

A distinguishing feature of such models is whether or not the gravitino is the lightest supersymmetric particle (LSP)\@.  Most analyses in the literature have focussed on the case of a heavy gravitino but, when the gravitino is the LSP, sparticles decay via the next-to-lightest supersymmetric particle (NLSP) and provide distinctive collider signals \cite{Kats:2011qh}.  In fact there is a broad class of models based on new gauge forces and extra dimensions (or locality) where supersymmetry breaking is transmitted to the first two generation sfermions in a flavour blind fashion, with third generation sfermions receiving suppressed contributions via gauge interactions~\cite{ArkaniHamed:1997fq, Dermisek:2006qj, Gabella:2007cp, Aharony:2010ch, Abel:2010vb, Craig:2011yk, Gherghetta:2011wc, Auzzi:2011eu, Csaki:2012fh, Delgado:2012rk, Larsen:2012rq, Craig:2012hc, Craig:2012di, Cohen:2012rm}.  Some of these models can simultaneously explain the fermion mass hierarchy~\cite{Gabella:2007cp,Craig:2011yk} and even incorporate unification~\cite{Craig:2012hc}  but, as of yet, are not strongly constrained by collider searches.  This motivates considering the phenomenology of natural supersymmetric models with a gravitino LSP, typical of models where supersymmetry breaking is mediated to the third generation sfermions by gauge interactions. These scenarios are collectively
referred to as natural gauge mediation (NGM).

We will study a particularly interesting subset of NGM models that contain a light bino, although our analysis also applies to other mediation mechanisms that yield a similar spectrum. Natural supersymmetry is usually constrained at the LHC by searches for jets plus missing energy from a neutralino LSP but, in NGM with a light bino, the neutralino can decay to a photon and gravitino.  The signal becomes two high energy photons plus missing energy, greatly reducing the Standard Model background, and was previously studied in \cite{Gabella:2007cp}. In this Letter we recast the 4.8 fb$^{-1}$ ATLAS diphoton search to find bounds on NGM models with a neutralino NLSP.\@  We investigate both electroweak and coloured sparticle production, producing robust and otherwise model independent bounds in the stop-gluino and neutralino-chargino mass planes respectively.  These limits provide the first nontrivial constraints on this class of models and allow us to quantify the degree of fine tuning required to avoid the bounds.

\section{Natural gauge mediation}

The NGM spectrum is determined by minimising the amount of fine tuning in the Higgs sector, retaining the features of gauge mediated supersymmetry breaking where possible. In particular we assume a light bino.  While this is not necessary from a naturalness point of view, a small value for $M_1$ is common in models with gauge mediated supersymmetry breaking.

In the minimal supersymmetric model there are three main sources of fine tuning~\cite{Kitano:2006gv}: the $\mu$-term
\be
\Delta_\mu=\frac{2\mu^2}{m_h^2}~,
\ee
one and two loop Higgs mass corrections from gauginos (with soft mass parameters $M_2$ and $M_3$)
\begin{align}
\Delta_2 & =\frac{3\alpha_2|M_2|^2}{2\pi m_h^2}\log{\pfrac{\Lambda}{\mbox{TeV}}}~, \\
\Delta_3 & =\frac{2y_t^2\alpha_3|M_3|^2}{\pi^3m_h^2}\frac{\tan^2\beta}{1+\tan^2\beta} \log^2{\pfrac{\Lambda}{\mbox{TeV}}}~,
\end{align}
and the one loop correction from the top Yukawa coupling ($y_t$)
\be
\Delta_t=\frac{3y_t^2(m_{\tilde{Q}_3}^2+m_{\tilde{u}_3}^2+|A_t|^2)}{4\pi^2m_h^2} \frac{\tan^2\beta}{1+\tan^2\beta}\log{\pfrac{\Lambda}{\mbox{TeV}}}~,
\ee
where $m_{\tilde{Q}_3}, m_{\tilde{u}_3}$ are the third generation soft mass parameters and $\Lambda$ represents the scale at which supersymmetry breaking is mediated.  For simplicity, we assume the decoupling limit such that $m_h$ denotes the physical Higgs boson mass and the mass scales in the logarithmns have been replaced by the TeV scale.  We will also neglect $A$-terms (i.e.\ $A_t$), which are generally predicted to be small for gauge mediated supersymmetry breaking.  Even if they are not (as occurs in some models~\cite{Dermisek:2006qj, Intriligator:2010be, Kang:2012ra, Craig:2012xp}) our conclusions are not affected as all bounds apply to the lightest, physical stop mass.

Requiring that the fine tuning is no worse than ten percent (i.e.\ $\Delta\lesssim10$) and assuming that $\Lambda$ is not much bigger than 10-100 TeV leads to the following constraints on the soft mass parameters:
\begin{align}
\label{ftbounds}
m_{\tilde{Q}_3},m_{\tilde{u}_3} & \lesssim 500-700\mbox{ GeV,} & M_3 & \lesssim 1100-2200\mbox{ GeV,}\\
\mu & \lesssim 300\mbox{ GeV,} & M_2 & \lesssim 1500-2000\mbox{ GeV.}
\end{align}
All sfermions other than the stops and the left handed sbottom can be decoupled since they do not significantly affect the fine tuning in the Higgs sector.\footnote{Sleptons only have a small effect on LHC phenomenology hence our results also apply to models where they are not decoupled.  The exception is when the NLSP is a slepton, typically a stau, which is common if the entire third generation is kept light.  Generalising stau searches \cite{ATLAS:2012ag, Aad:2012rt, CMS-PAS-SUS-12-004} to include light stops is therefore a well motivated extension to this work.}  Hence a common scale will be assumed for the remaining soft mass parameters
\be\label{eq:mhigh}
m_{\tilde{Q}_{1,2}},\,m_{\tilde{u}_{1,2}}, m_{\tilde{d}_{1,2,3}},\,m_{\tilde{L}_{1,2,3}},\,m_{\tilde{e}_{1,2,3}},\,m_{\tilde{\nu}_{1,2,3}}
\equiv\widetilde{M}~,
\ee
where $\widetilde{M}>$ few TeV\@. This leaves the minimal sfermion spectrum required for naturalness.

In the gaugino sector we first assume that the standard one loop relationship
\be\label{eq:NGMgaugino}
\frac{3}{5} \frac{M_1}{\alpha_1}=\frac{M_2}{\alpha_2}=\frac{M_3}{\alpha_3}~,
\ee
i.e.\ gaugino unification, continues to hold. Combined with the existing gluino mass bound from naturalness, this implies that
\be
M_1=\frac{5}{3}\frac{\alpha_1(M_Z)}{\alpha_3(M_Z)}M_3\lesssim160-320\mbox{ GeV}
\ee
which will be important in determining the identity of the NLSP\@.  In the Higgs sector we fix $m_h=125$ GeV\@.  A number of examples with an NGM spectrum~\cite{Craig:2011yk, Auzzi:2011eu, Csaki:2012fh, Larsen:2012rq, Craig:2012hc} already contain a mechanism to raise the Higgs mass. However, it should be stressed that we only consider the fine tuning of a minimal supersymmetric model with a $\mu$-term, and that any mechanism which raises the Higgs mass may change the amount of tuning.

\begin{figure}[!tb]
\begin{center}
\includegraphics[width=0.8\textwidth]{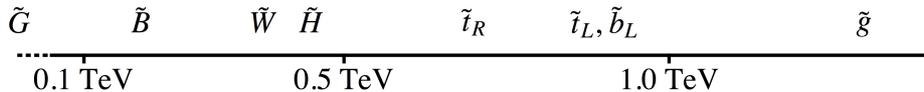}
\caption{A typical spectrum for natural gauge mediation.\label{fig:NGMspectra}}
\end{center}
\end{figure}

Little parameter space remains for stop or sbottom NLSPs\@.  Searches for direct stop production \cite{ATLAS-CONF-2012-074, ATLAS-CONF-2012-073, ATLAS-CONF-2012-070, ATLAS-CONF-2012-059, ATLAS-CONF-2012-071} exclude stop masses between 220 and 500 GeV, and between 110 and 165 GeV\@.  In addition, left handed stop or sbottom NLSPs below 350 GeV and a right handed stop NLSP below about 200 GeV were excluded in Ref.~\cite{Papucci:2011wy} (although a stop NLSP close to the top mass is not yet excluded).  Since eq.~\eqref{eq:NGMgaugino} implies that $M_1<M_2<M_3$ there are two further possibilities.  The first is a Higgsino-rich chargino, requiring $|\mu|<|M_1|\lesssim 160-320$ GeV\@.  This scenario turns out to be somewhat non-generic \cite{Kribs:2008hq} so we do not focus on it here (although chargino NLSPs outside of an NGM context have been investigated in Ref.~\cite{Chatrchyan:2011ah}).  By far the most common NLSP is a bino or Higgsino-rich neutralino, also lighter than about $160-320$ GeV\@.  NGM models with a Higgsino-rich neutralino NLSP decaying to $Z$ bosons were analysed in Ref.~\cite{Aad:2012cz}, so we focus on bino-rich neutralinos, which decay to photons and gravitinos.  A typical spectrum is given in Figure \ref{fig:NGMspectra}.

Relaxing the constraint imposed by eq.~\eqref{eq:NGMgaugino} the spectrum may remain similar or may be quite different.  For example, it is well known that many explicit models only generate gaugino masses at two loops.  This further suppresses $M_1$ so the NLSP is still a bino-rich neutralino.  On the other hand, one can easily construct more involved models where the gaugino mass hierarchy implied by eq.~\eqref{eq:NGMgaugino} is completely disrupted, whereupon alternative NLSPs become possible.  To constrain such models one would need to repeat the relevant analysis in Ref.~\cite{Kats:2011qh}, but with a light third generation.  Generically one would expect to find stronger bounds on gluino and NLSP masses, but weaker bounds for the stop mass relative to those found for degenerate squarks.

In order to constrain the NGM framework we take a purely phenomenological approach using simplified models.  This enables robust bounds to be placed on physical masses independently of model details.  Since fine tuning requirements force all NGM models to contain light Higgsinos, light stops and light gluinos, one expects both coloured and electroweak sparticle production.  We consider separate simplified models for each process.  In practice both processes contribute in any NGM model, hence our bounds are always conservative.

For coloured production we decouple all sparticles other than the gluino, the right-handed stop and the bino.  Hence sparticle creation proceeds through gluino or stop pair production.  Including other light sparticles would strengthen our final bounds.  The bino must be light enough such that the NLSP is mostly bino (therefore decays photons) but otherwise its mass has only a weak kinematic effect on the signal strength.  This can be seen in Refs.~\cite{ATLAS-CONF-2012-072, CMS-PAS-SUS-12-018} and we have verified that the situation is the same here.  We thus fix the bino mass at 100 GeV in this scenario.  For electroweak production we decouple all coloured sparticles and keep only the Higgsinos and bino in the electroweak sector.  Sparticle creation now proceeds through neutralino/chargino pair production.  Naturalness does not forbid heavy winos and including them among the light states again leads to stronger bounds.

\section{Exclusion limits}

Neutralino NLSPs decaying to photons lead to a very distinctive signature: two high energy photons and missing transverse momentum from the gravitinos. This channel has a low background and is the subject of recent updates from the ATLAS \cite{ATLAS-CONF-2012-072} and CMS \cite{CMS-PAS-SUS-12-018} experiments. Only the ATLAS diphoton search is considered here, on the basis that CMS and ATLAS have a similar reach. We also considered the earlier, 1 fb$^{-1}$ ATLAS search \cite{Aad:2011zj}, which has softer kinematic requirements, but found that it offered no additional constraints on NGM.

Three signal regions were defined in the ATLAS diphoton analysis. Of these the first two are the most constraining for coloured production and the third for electroweak production. At least two isolated photons with $p_T >$ 50 GeV are required, for which the energy in a cone of $\Delta R=0.2$ surrounding the photon's deposition in the calorimeter must be less than 5 GeV\@. Table~\ref{tab:cuts} summarises the further selection cuts for each signal region, the results of the ATLAS search, and the ATLAS derived limits on the number of events attributable to new physics. These results may be used to constrain the NGM parameter space if we evaluate the signal expectation for candidate NGM models.

We use \tt SUSY-HIT 1.3 \rm\cite{Djouadi:2006bz} to produce mass and decay spectra for a given set of model parameters, \tt Pythia 6.24.26 \rm\cite{Sjostrand:2006za} to generate 10,000 Monte Carlo events for each point, and a custom version of \tt Delphes 1.9 \rm\cite{Ovyn:2009tx} (with photon isolation added) to provide a fast ATLAS detector simulation. The total supersymmetric production cross section is calculated at next-to-leading order using \tt PROSPINO 2.1 \rm\cite{Beenakker:1996ed}. A cone-based overlap removal procedure is used to avoid double counting particles that are reconstructed as more than one object (e.g.\ an electron and a jet).

\begin{table}[!t]
\begin{center}
\begin{tabular}{|l|c|c|c|c|c|}
\hline
Region & R1 & R2 & R3 \\
\hline
Photons ($E_T>50$ GeV)&$\ge 2$&$\ge 2$&$\ge 2$\\
$E_T^\mathrm{miss}$ (GeV) & $>200$&$>100$ &$>125$ \\
$\Delta\phi(\gamma, p_{T}^\mathrm{miss})_\mathrm{min}$ & $>0.5$ & $-$ & $>0.5$ \\
$H_T$ (GeV) & $>600$ &  $>1100$  &  $-$  \\
\hline
Expected background & 0.10 $\pm$ 0.03 $\pm$ 0.07 & 0.36 $\pm$ 0.05 $\pm$ 0.27 & 2.11 $\pm$ 0.37 $\pm$ 0.77 \\
Observed events & 0 & 0 & 2 \\
\hline
95\% CL upper limit & 3.1 & 3.1 & 4.9 \\
\hline
\end{tabular}
\end{center}
\caption{Selection cuts for the three ATLAS diphoton search signal regions. $\Delta\phi(\gamma, p_{T}^\mathrm{miss})_\mathrm{min}$ is the smallest of the azimuthal separations between the missing momentum $p_{T}^\mathrm{miss}$ and the momenta of the two leading photons in the event. The total visible transverse energy $H_T$ is the scalar sum of the transverse momenta of the jets, leptons and two leading photons in the event.\label{tab:cuts}}
\end{table}

\begin{figure}[!tb]
\begin{center}
\includegraphics[width=0.55\textwidth]{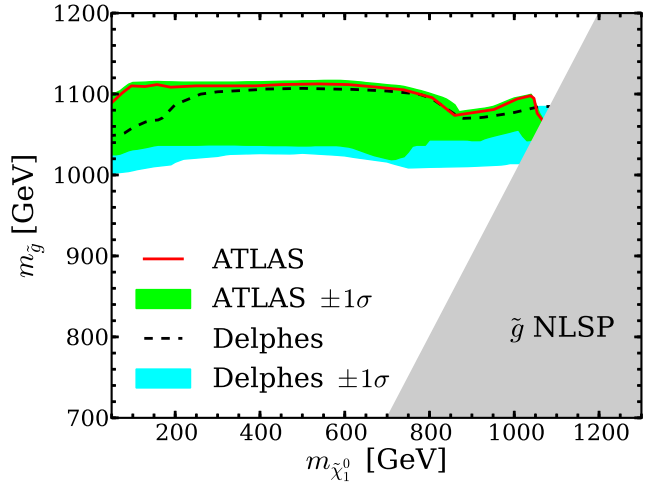}
\caption{Comparison between the \tt Delphes \rm and ATLAS 95\% exclusion limits in the $m_{\tilde{\chi}_1^0}-m_{\tilde{g}}$ plane. The \tt Delphes \rm limit is obtained by taking the union of the \tt Delphes \rm limits for each signal region, with the limits tuned to match the ATLAS results.\label{fig:AtlasLimits}}
\end{center}
\end{figure}

To approximate the ATLAS limit setting procedure we follow the approach used in Ref.~\cite{Allanach:2011wi}. We use the published ATLAS limit in the $m_{\tilde{\chi}_1^0}-m_{\tilde{g}}$ plane to calibrate systematic error parameters in a simplified model of the ATLAS likelihood function, then use this likelihood function to generate limits in the NGM parameter space. We have confirmed that our procedure reproduces the ATLAS 95\% confidence level exclusion contours and, where modest discrepancies are encountered, we tune our parameters to ensure that our results are more conservative.  See Figure \ref{fig:AtlasLimits} for details.

In our exploration of the NGM parameter space, we fix $\tan\beta=2$, ${\widetilde M}=2.5$ TeV and $c\tau_{\tilde{\chi}_1^0}<0.1$ mm (to ensure prompt neutralino decays) throughout.  For coloured production we fix $m_{\tilde{Q}_3}$, $\mu$ and $M_2$ at the high scale ${\widetilde M}$, $M_1=100$ GeV, then scan over $m_{\tilde{u}_3}$ (approximately equal to $m_{\tilde{t}_1}$ when $m_{\tilde{u}_3}>m_t$) and $M_3$ (i.e.\ the gluino mass $m_{\tilde{g}}$).  For electroweak production we instead fix $m_{\tilde{Q}_3}$, $m_{\tilde{u}_3}$, $M_2$ and $M_3$ at the high scale ${\widetilde M}$, then scan over $M_1$ (approximately equal to $m_{\tilde{\chi}^0_1}$ when $M_1<\mu$) and $\mu$ (approximately equal to $m_{\tilde{\chi}^\pm_1}$ when $M_1<\mu$).

\begin{figure}[!tb]
\begin{center}
\includegraphics[width=1.0\textwidth]{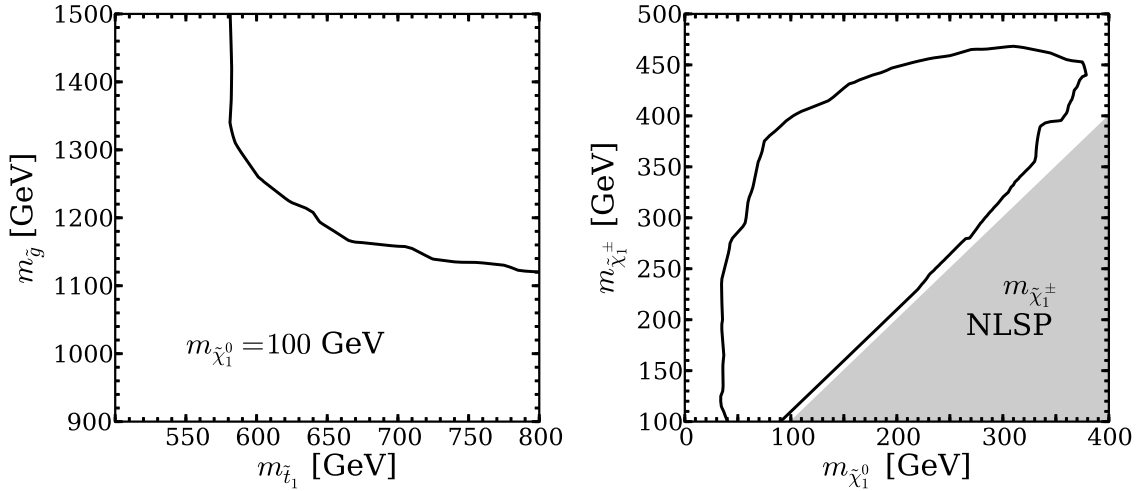}
\caption{The 95\% confidence level exclusion contours.  {\em Left}: Limits on coloured production in the $m_{\tilde{t}_1}-m_{\tilde{g}}$ plane (note that the limit does not degrade at low stop mass for $m_{\tilde{\chi}_1^0}=100$~GeV\@).  {\em Right}: Limits on electroweak production in the $m_{\tilde{\chi}_1^{0}}-m_{\tilde{\chi}_1^{\pm}}$ plane.  Both production processes are active in any given model so all limits are conservative.\label{fig:colLimits}}
\end{center}
\end{figure}

The resulting limits are shown in Figure~\ref{fig:colLimits}.  For coloured production the gluino mass limit approaches $\approx1100$~GeV for heavy stops, in agreement with the published limits from ATLAS and CMS \cite{ATLAS-CONF-2012-072, CMS-PAS-SUS-12-018} (note that varying the bino mass does not significantly change the limits as can be seen from Figure~\ref{fig:AtlasLimits}). Consequently one must accept a fine tuning of at least $10-40$\% due to two loop Higgs mass corrections from gluinos.  Stop masses less than $\approx 580$ GeV are excluded (the effect of including weak production as well would raise this limit further), corresponding to a fine tuning of at least $8-17$\% due to one loop Higgs mass corrections from the top Yukawa coupling.  These limits do not degrade at low stop mass as photons from neutralino decays typically remain hard enough to pass all cuts for $m_{\tilde{\chi}^0_1}=100$ GeV\@.

Limits on weak production reveal that a significant slice of the $m_{\tilde{\chi}_1^{0}}-m_{\tilde{\chi}_1^{\pm}}$ plane is also excluded.  The limit does not extend to neutralino masses below 50 GeV as too few events in these models pass the missing energy requirement of the ATLAS search.  Nor does it extend to $m_{\tilde{\chi}_1^{0}}\approx m_{\tilde{\chi}_1^{\pm}}$ where the branching ratio of neutralinos to photons drops off.  One could in principle have models with $m_{\tilde{\chi}_1^{0}}<50$ GeV and a small value of $\mu$ (and hence low fine tuning) although, if one insists on gaugino unification, this is difficult to achieve due to limits on the gluino mass from coloured production.  If $m_{\tilde{\chi}_1^{0}}>50$ GeV one must have $m_{\tilde{\chi}_1^{\pm}}\gtrsim300-470$ GeV and a fine tuning of at least $4-9$\% from the $\mu$-term.

Combining both limits one can search for the most natural NGM model (with a neutralino NLSP) not yet excluded by the LHC\@.  If gaugino unification is assumed both charginos, from the Higgsino and wino, are light.  Constraints from electroweak production are therefore severe.  Choosing
\begin{align}
m_{\tilde{g}} & \gtrsim2200\mbox{ GeV,} & \mu & \gtrsim1000\mbox{ GeV,}
\end{align}
avoids both limits and yields a fine tuning of at least 1\% from the $\mu$-term.  The fine tuning from the top Yukawa coupling is subdominant.  The constraints are relaxed if gaugino unification is not assumed such that the wino is allowed to be heavy and the bino light.  Then
\begin{align}
m_{\tilde{g}} & \gtrsim1600\mbox{ GeV,} & \mu & \gtrsim400\mbox{ GeV,} & m_{\tilde{Q}_3},m_{\tilde{u}_3} & \gtrsim800\mbox{ GeV,}
\end{align}
avoids both limits with a fine tuning of at least 5\% from the $\mu$-term and $5-9$\% from the top Yukawa coupling.  As expected, the fine tuning in a realistic spectrum is worse than that suggested by the simplified models alone.

Finally, we note that the current ATLAS search has not been optimised for direct stop production. A more dedicated search using $b$-tagging and reoptimised selections on $E_T^{\rm miss}$ and $H_T$ could increase the reach, and we strongly encourage effort in this area.

\section{Conclusion}
In summary, using a 4.8 fb$^{-1}$ ATLAS diphoton search, we have placed mass limits on simplified models of natural gauge mediation with a neutralino NLSP for coloured and electroweak sparticle production.  Because both production mechanisms are active in actual realisations of natural gauge mediation, our bounds are conservative.  Top squarks with a mass below $\approx580$ GeV are excluded, as are gluinos with a mass below $\approx1100$ GeV.  Assuming the neutralino is heavier than 50 GeV means that charginos lighter than $\approx 300-470$ GeV are ruled out.  Otherwise a neutralino lighter than $\approx50$ GeV may allow our bounds to be evaded.  This places the first nontrivial constraints that test the naturalness of this class of models.

\section*{Acknowledgements}

This work was supported by the Australian Research Council. MJW is supported by ARC Discovery Project DP1095099. TG thanks the SITP at Stanford and CERN TH division for hospitality during the completion of this work. This material is based upon work supported in part by the National Science Foundation under Grant No.1066293 and the hospitality of the Aspen Center for Physics.

\bibliographystyle{JHEP-2}
\bibliography{bib}
\end{document}